\documentclass[preprint2]{aastex62}

\usepackage{amsmath}
\usepackage{graphicx}
\usepackage[T1]{fontenc}
\usepackage{booktabs}
\usepackage{lipsum}
\usepackage{txfonts} 
\usepackage[figure,figure*]{hypcap} 
\usepackage{chngcntr}

\begin{document}

\title{What Is Inside Matters: Simulated Green Valley Galaxies Have Too Centrally Concentrated Star Formation}

\shorttitle{Too Centrally Concentrated SFRs in Simulated Galaxies}
\shortauthors{Starkenburg et al.}

\author{Tjitske K. Starkenburg}
\affil{Flatiron Institute, 162 5th Avenue, New York NY 10010, USA}
\author{Stephanie Tonnesen}
\affil{Flatiron Institute, 162 5th Avenue, New York NY 10010, USA}
\author{Claire Kopenhafer}
\affil{Michigan State University, 567 Wilson Road, East Lansing MI 48824, USA}
\correspondingauthor{Tjitske Starkenburg}
\email{tstarkenburg@flatironinstitute.org}

\begin{abstract}
In spatially resolved galaxy observations, star formation rate radial profiles are found to correlate with total specific star formation rates. A central depletion in star formation is thought to correlate with the globally depressed star formation rates of, for example, galaxies within the Green Valley. We present, for the first time, radial specific star formation rate profiles for a statistical sample of simulated galaxies from the Illustris and EAGLE large cosmological simulations. For galaxies on the star-forming sequence, simulated specific star formation rate profiles are in reasonable agreement with observations.

However, both galaxy samples show centrally concentrated star formation for galaxies in the Green Valley at all galaxy stellar masses, suggesting that quenching occurs from the outside-in, in strong conflict with observations of inside-out quenching.
This difference between simulations and observations may be due to resolution issues and/or possible failures in the star formation and feedback implementation in current large-scale cosmological simulations. We conclude that the distribution of star formation within galaxies is a strong additional constraint for simulations and models, in particular related to the quenching of star formation.
\end{abstract}

\section{Introduction}\label{intro}

How galaxies stop forming stars, i.e. quench, is an outstanding question in both observations and simulations \citep[e.g.][]{ManBelli2018}. Observations of the star formation rates (SFR) and stellar masses of galaxies find that the star-forming galaxy population tends to lie along a star-forming sequence (SFS) that depends on galaxy stellar mass, with a large tail of galaxies with low SFRs \citep[e.g.][]{Noeske2007, Daddi2007, Salim2007}. This low-SFR tail is often called the green valley (GV), through which galaxies pass from blue and star-forming to red and quenched \citep[e.g.][]{Faber07, Martin2007, Schiminovich2007, peng2010}.

Reproducing the distribution of galaxies in the SFR--stellar mass plane is one of the requirements of a successful large-scale cosmological simulation.  In massive halos, shock-heated gas may not form stars
 \citep[e.g.][]{Silk1977, Rees1977, Binney1977, Birnboim2003}, and satellite galaxies can be quenched due to gas removal or consumption \citep[e.g.][]{Boselli2006}; yet in order to match the observed fraction of quenched galaxies, simulations must include feedback from both supernovae and active galactic nuclei \citep[e.g.][and references therein]{somerville2015}.  

Although current large-scale simulations include a diversity of feedback prescriptions, they all qualitatively reproduce the general distribution of galaxies on the SFR--mass plane (e.g. \citealt{Vogelsberger14,Genel14,schaye2015}, see \citealt{somerville2015,Hahn2018} for a comparison). However, recent observations have radially mapped the SFR in galaxies, providing a more stringent test of star formation and quenching in galaxies. For example, using
the CALIFA survey, \citet{Perez13} find that the central regions of massive galaxies (log(M$_*$/M$_{\sun}$) > 10.5) are older than the outer disks, while at lower masses the age gradient flattens. Similar trends with regards to stellar mass are found in color gradients \citep{Pan2015}, while \citet{Ibarra-Medel2016} find a larger age gradient diversity for lower mass systems.

The MaNGA survey \citep{Bundy2015,Yan2016,Blanton2017} is obtaining spatially resolved spectroscopy for nearby galaxies. Using this survey, \citet[][hereafter B18]{Belfiore18} present the radial profiles for galaxies on and above the SFS and for galaxies below the SFS (GV galaxies) for a range of stellar masses. In general they find that with increasing stellar mass, the specific SFR (SFR/M$_*$, sSFR) in the central regions of galaxies is more depressed relative to that beyond one half-light radius ($1 R_e$). For galaxies with log(M$_*$/M$_{\sun}$) > 10, the difference between the sSFR in the central and outer regions is significantly more pronounced in GV galaxies than in star-forming galaxies, reaching differences ${\gtrsim}1$~dex. 

Similar central sSFR suppression was found for both the CALIFA and SAMI surveys, increasing toward earlier galaxy Hubble type \citep{GonzalezDelgado2016, Medling2018}. \citet{Nelson2016} find a correlation between a galaxy's distance from the SFS and the SFR at all radii in stacked H$\alpha$ maps for galaxies at $z\sim1$, but see additional central suppression for galaxies with ($10.5 < \textrm{log}(M_*/M_{\sun}) < 11$). \citet{Ellison2018} use the individual spaxels from the MANGA survey and find that the resolved local SFR is increased or depressed at all radii in correlation with the total SFR. For galaxies with total SFRs more than 1~dex below the SFS the SFR shows additional depression in the central regions.

Zoom cosmological simulations have found central star formation enhancement and depletion for high-redshift galaxies \citep{Tacchella2016}. Additionally, \citet{Orr2017} find that stacked radial profiles correlate with total sSFR for a sample of galaxies while individual galaxies show strong variability over time in their SFR profiles. However, these zoom cosmological simulations lack AGN feedback and are unable to fully model the quenching of galaxies. Moreover, no results for (s)SFR profiles exist for statistical samples across a large range in stellar masses and total sSFR.

In this letter we examine the radial dependence of sSFR in large-scale cosmological simulations in order to compare to observations, specifically B18. We use the publicly available data from cosmological simulations using different hydrodynamic solvers: the Illustris \citep{Vogelsberger14,Genel14} and EAGLE \citep{schaye2015,crain2015} simulations.  These simulations also use different stellar and AGN feedback prescriptions; which also differ from IllustrisTNG \citep{Pillepich2018}, recently publicly released. 

In Section \ref{methods} we briefly introduce the Illustris and EAGLE simulations. We present our radial profiles in Section \ref{results}, and discuss these in Section \ref{discussion}. We summarize our conclusions in Section \ref{conclusions}.

\begin{figure}
\includegraphics[width = .5\textwidth]{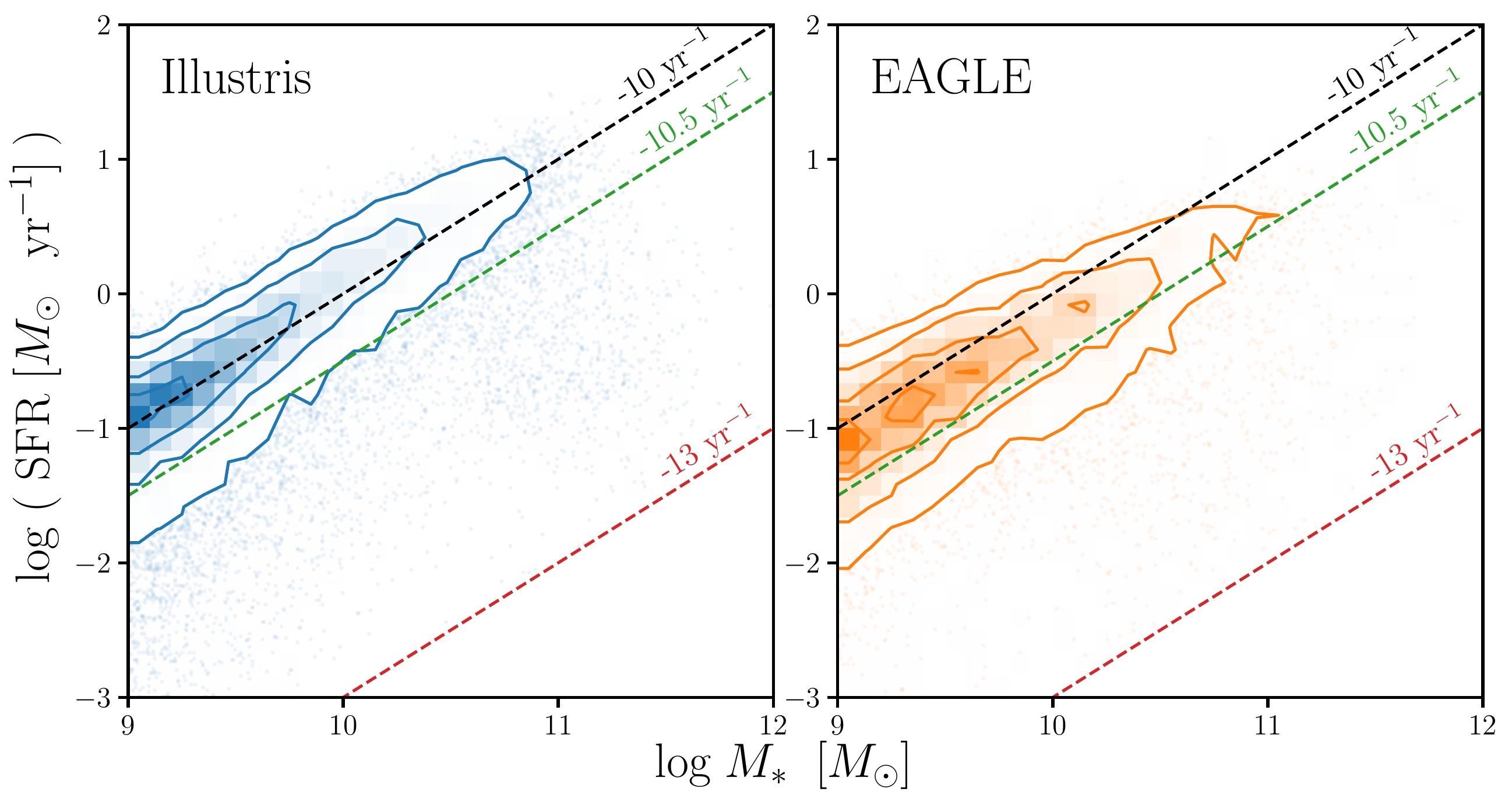}
\caption{\label{figs:SFMS} The total SFR versus total stellar mass distribution for all galaxies in our samples ($9 < \textrm{log}(M_*/M_{\sun}) < 12$) from the Illustris (left) and EAGLE (right) simulations, where both $M_*$ and SFR are summed within 3Rhalf. 
For reference, dashed lines indicate log(sSFR~yr)~$ = -10$ (black), log(sSFR~yr)~$ = -10.5$ (green), and log(sSFR~yr)~$ = -13$ (red).}
\end{figure}

\begin{figure*}
\includegraphics[width = \textwidth]{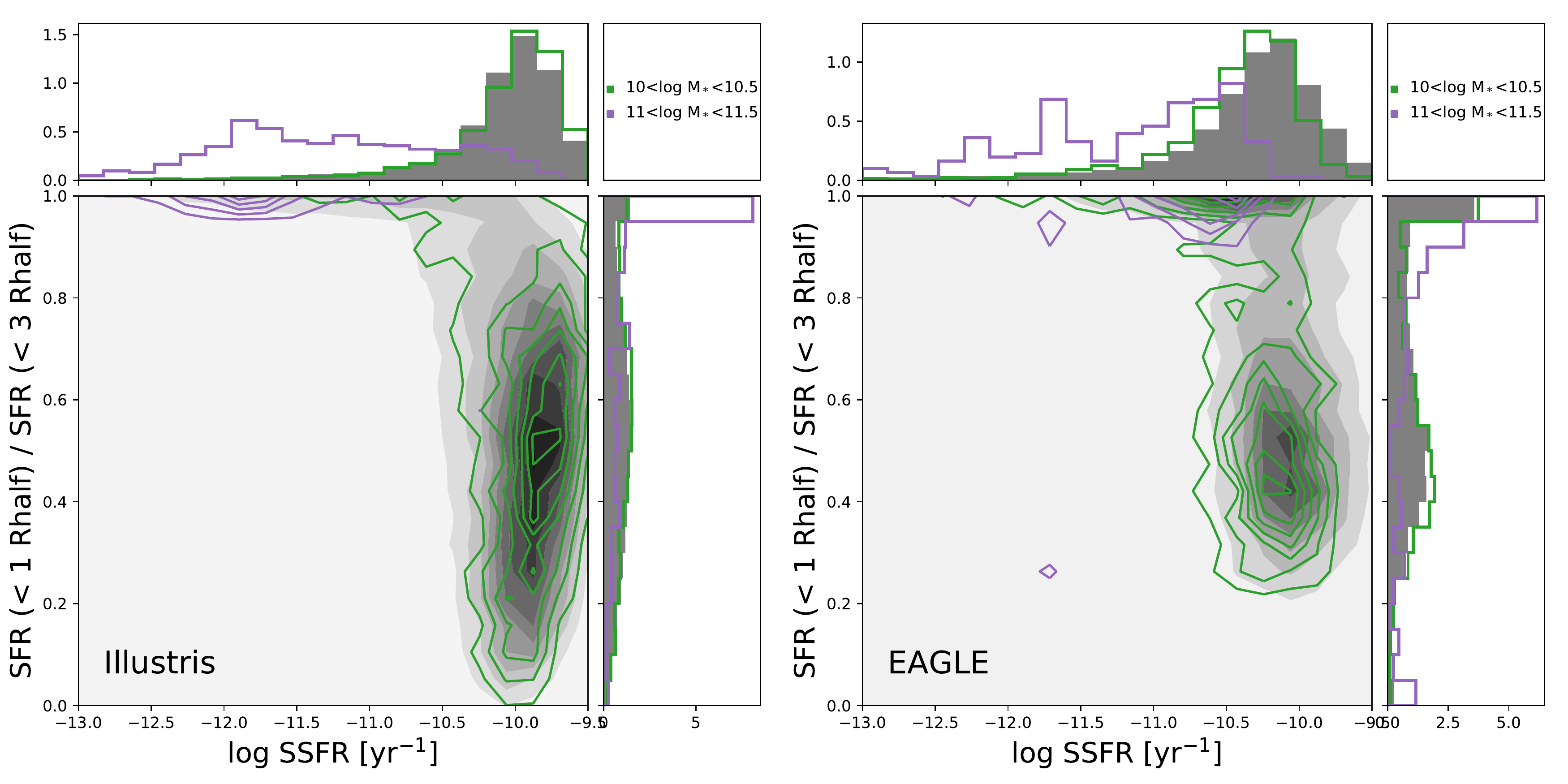}
\caption{\label{figs:contours} SFR$_{\rm 1Rhalf}$/SFR$_{\rm 3Rhalf}$ versus sSFR$_{\rm 3Rhalf}$ for all galaxies in our samples (gray contours) from the Illustris (left) and EAGLE (right) simulations. Two subsets of galaxies are highlighted: $10.0 < \textrm{log}(M_*/M_{\sun}) < 10.5$ (green) and $11.0 < \textrm{log}(M_*/M_{\sun}) < 11.5$ (purple). The SFR in most star-forming galaxies is more centrally concentrated in EAGLE than in Illustris.}
\end{figure*}

\section{Data}\label{methods}
Our sample consists of $z=0$ galaxies from the Illustris and EAGLE simulations with stellar masses within the mass range in B18: $9 < \textrm{log}(M_*/M_{\sun}) < 12$. 

The Illustris simulation\footnote{\url{http://www.illustris-project.org}} \citep{Vogelsberger14,Genel14} is a cosmological hydrodynamic simulation with a ($106$Mpc)$^3$ volume using the {\sc Arepo} moving-mesh code \citep{springel2010} with a uniform baryonic mass resolution of $1.26 \times 10^6\ M_{\sun}$, and gravitational softening length of $0.7$~kpc for collisionless baryonic particles at $z \lesssim 1$. This value is also the minimum gravitational softening for gas, which is tied to the cell size. Subgrid models implement star formation and feedback \citep{springel2003}, and black hole accretion and dual-mode AGN feedback \citep{sijacki2007, vogelsberger2013}.  We use the public data for Illustris-1 \citep{Nelson2015}.

The Evolution and Assembly of GaLaxies and their Environment (EAGLE)\footnote{\url{http://www.eaglesim.org}} project of the Virgo Consortium \citep{schaye2015, crain2015} consists of a suite of cosmological hydrodynamic simulations, run using a modified version of the N-body/SPH code Gadget3 \citep[lastly described in][]{springel2005} called {\sc Anarchy} (Dalla Vecchia et al. in prep.; see also \citealt{schaye2015,schaller2015}), with sub-grid models for star formation \citep{schaye2008}, black hole formation, accretion, and AGN feedback \citep{booth2009, schaye2015}. The baryonic particle resolution is $1.81 \times 10^6\ M_{\sun}$ and the gravitational softening is $0.7$~kpc at $z < 2.8$. We use the large box reference simulation of the EAGLE suite (RefL0100N1504) with a volume of ($100$ Mpc)$^3$, and use the public data release \citep{McAlpine2016}.

In this work we define the total stellar mass, SFR, and sSFR as the total within 3 times the stellar half mass radius (3Rhalf), although our results do not depend on the maximum radius used. We choose to include star formation at larger radii than B18 (maximum radius of $2.5 R_e$) in order to make sure that we include galaxies with no central star formation out to large radii \citep[for spheroidal galaxies $2.5 R_e$ approximates 3Rhalf; see e.g. the discussion in][]{somerville2018}.  We examine the sSFR profiles of currently star-forming galaxies, and therefore include galaxies with total log(sSFR~yr$^{-1}$)~$>-13$. We require galaxies in our samples to have Rhalf > 4 kpc, which is slightly larger than the size of one MANGA fiber, but ensures that Rhalf is well resolved.

\section{Results}\label{results}

Figure~\ref{figs:SFMS} shows the total SFR--stellar mass relation for our samples. Importantly, both simulations form a well-defined star-forming sequence. 
The Illustris SFS has log(sSFR~yr$^{-1}$) slightly above $-10$ while the EAGLE SFS is below $-10$. Because of this difference, we tested multiple definitions for the GV: as galaxies with SFR $< 0.39$~dex below the SFS and log(sSFR~yr$^{-1}$)~$> -13$ (following B18), and as galaxies with $-13 <$~log(sSFR~yr$^{-1})$~ $< -10.5$ or $-13 <$~log(sSFR~yr$^{-1}$)~$< -11$.  The linear fitted SFS for Illustris and EAGLE are log(SFR)~$= m (\textrm{log}(M_*)-10.5)+b$ with $m=1.01$ and $b=0.59$ for Illustris, and $m=0.91$ and $b=0.23$ for EAGLE \citep{Hahn2018}. Qualitatively, our results do not depend on the GV definition, so for consistency across the two simulations and the observations, we follow the first GV definition. 
Our Star-Forming samples consists of all galaxies with SFR $>$ SFS $-0.39$~dex.

The central concentration of the SFR in the galaxies is illustrated in Figure~\ref{figs:contours}. The gray contours show our complete sample from $9 < \textrm{log}(M_*/M_{\sun}) <12$.  To illustrate any mass dependence of this relation, we overplot the contours of two mass bins as green ($10 < \textrm{log}(M_*/M_{\sun}) < 10.5$) and purple ($11 < \textrm{log}(M_*/M_{\sun}) < 11.5$).  As expected, in both simulations we find that more massive galaxies tend to have lower total sSFRs.  More massive galaxies also tend to have more centrally concentrated SF.  In  Illustris (left panel), strongly star-forming galaxies generally do not have centrally-concentrated SF, but as the sSFR decreases, the SF becomes more centrally concentrated. In EAGLE (right panel), the distribution of SF in star-forming galaxies is more centrally concentrated than in Illustris. 

Nevertheless, at low sSFRs (log(sSFR~yr)$_{\rm 3Rhalf} < -10.5$) galaxies in both simulations have strongly centralized SF.

\begin{figure*}
\includegraphics[width = \textwidth]{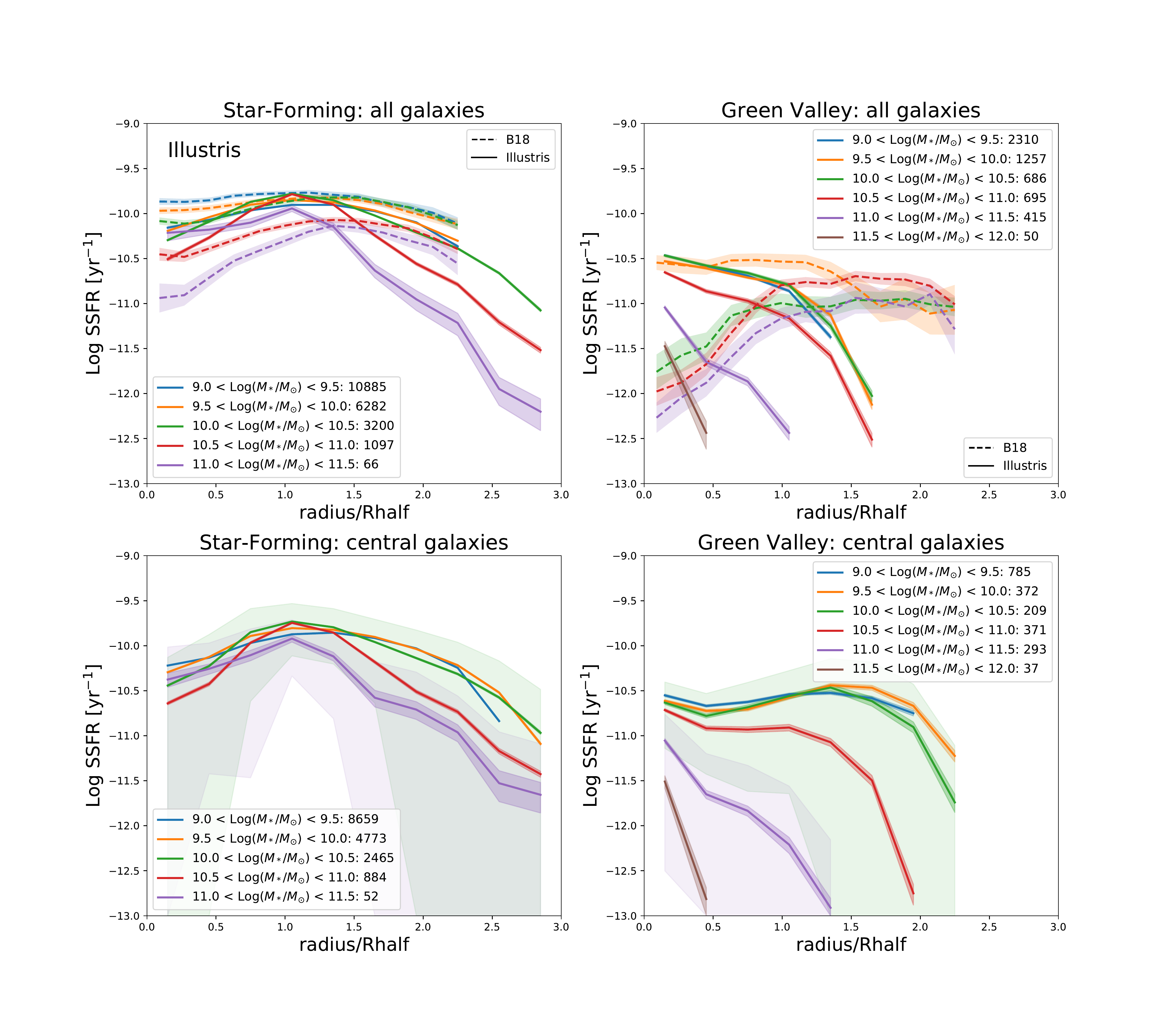}
\caption{\label{figs:radIllustris} The log(sSFR) profiles for our Illustris Star-Forming (left) and GV (right) samples, including all galaxies (top panels), or only central galaxies (bottom panels). Lines show the log(sSFR) profile for galaxies binned by their total stellar mass. The profiles are determined by the Tukey biweight (the same estimator as in B18). The light shaded regions show the robust biweight scale estimator for two mass bins, while the darker shaded regions show this scale estimator divided by $\sqrt{N}$ (following B18). Using the median for the log(sSFR) profiles
results in the same trends. Data from B18 are overplotted in the top panels (dashed lines). Like B18, mass bins with $\ge$ 20 galaxies are shown.}
\end{figure*}
\begin{figure*}
\includegraphics[width = \textwidth]{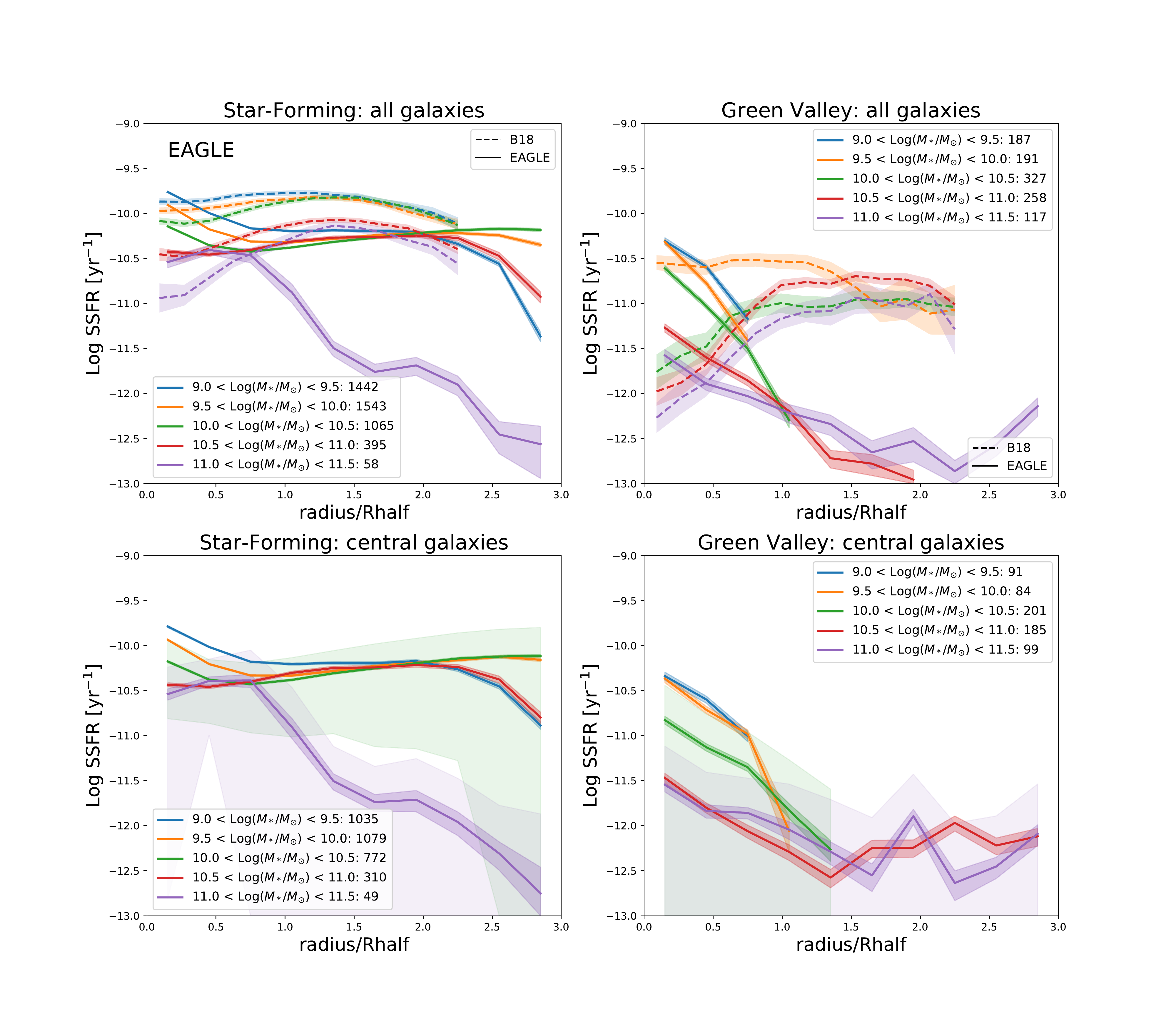}
\caption{\label{figs:radEAGLE} As Figure~\ref{figs:radIllustris}, for our EAGLE sample.}
\end{figure*}
\begin{figure*}
\includegraphics[width = \textwidth]{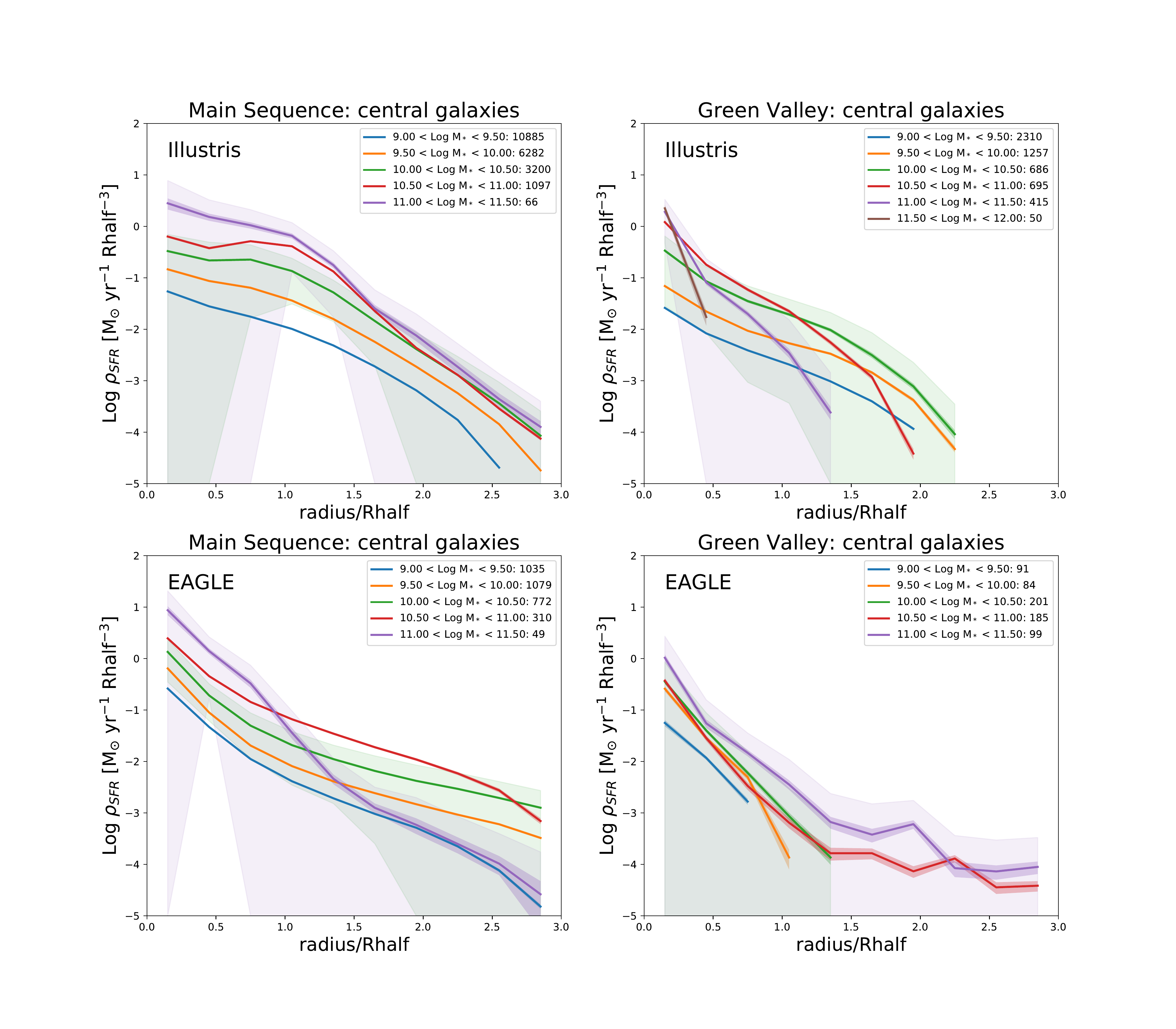}
\caption{\label{figs:SFR} Log~$\rho_{SFR}$ profiles for Star-Forming (left), and Green Valley (right) central galaxies in Illustris (top) and EAGLE (bottom).}
\end{figure*}
In order to best compare with B18, we plot radial profiles of sSFR in these simulated galaxies in Figures \ref{figs:radIllustris} and \ref{figs:radEAGLE}, using the B18 mass bins. We measure sSFR volume density in radial bins
of $0.3$Rhalf (our results are insensitive to differences
between sSFR volume density ($\rho_{\rm SFR}/\rho_{M_*}$) and sSFR
surface density ($\Sigma_{\rm SFR}/\Sigma_{M_*}$), so we use the more physically meaningful volume density). These bins are not well resolved for all galaxies, an effect we discuss in Section \ref{discussion}. As the star formation distribution in satellite galaxies can be affected by environmental effects we show sSFR profiles considering all galaxies (top panels) and only central galaxies (bottom panels). 

For comparison we show the B18 data in the top panels, using a correction of Rhalf~$= 1.2 R_e$~\citep{somerville2018}.

We first focus on Illustris galaxies (Figure \ref{figs:radIllustris}).  For the Star-Forming sample (left panels), for all mass bins, the log(sSFR~yr$^{-1}$) is near $-10.5$ in the center,
then increases slightly with radius (${\sim}0.5$~dex) to peak around 1--1.5Rhalf, and drops off toward larger radii, where the drop off is larger for higher mass galaxies. This pattern is almost identical when considering either all galaxies or only central galaxies, and is in reasonable agreement with B18 until radii $\ge$1.5Rhalf, where B18 find flatter profiles.  

For GV galaxies the central sSFR volume density depends strongly on mass with values around $-10.5$ for galaxies with $9 < \textrm{log}(M_*/M_{\sun}) < 11$, and values ${\sim}0.5$--$1$~dex lower for more massive galaxies. 

In all cases the sSFR drops sharply at larger radii. 
For galaxies with $\textrm{log}(M_*/M_{\sun}) < 10.5$, the star formation profile is significantly flatter within $1.5$Rhalf and more extended for central galaxies than for the whole sample of GV galaxies. Only the $9.5 < \textrm{log}(M_*/M_{\sun}) < 10$ mass bin of central GV galaxies in Illustris agrees well with the B18 data. 
For all more massive systems the B18 lines show strong central depletion in sSFRs. Even considering the significant scatter, shown in the lighter shaded bands, \emph{the decreasing sSFR with increasing radius in Illustris GV galaxies is strikingly different from the 
low central sSFRs observed in B18}.

Figure \ref{figs:radEAGLE} shows the sSFR volume density of galaxies from the EAGLE simulation.  Star-Forming galaxies have a more centrally concentrated sSFR density than those in Illustris, as is also seen in Figure~\ref{figs:contours}.  
For all galaxies with $\textrm{log}(M_*/M_{\sun}) > 9.5$, the sSFR profiles have similar values, and are flat from $\le$~1Rhalf to 3Rhalf, in contrast to the steeply decreasing profiles (beyond ${\sim}1$Rhalf) in the Illustris galaxy population. As in the Illustris sample, the difference between the sSFR profiles for all Star-Forming galaxies and only central Star-Forming galaxies is minimal. The sSFR profiles are in reasonable agreement with B18 at larger radii (except for the lowest and highest mass bins), but almost all mass bins show additional star formation in the centers. The exception is the $10.5 < \textrm{log}(M_*/M_{\sun}) < 11.0$ bin: the sSFR profile of this mass bin matches well with the observed sample in B18. 

In the EAGLE GV population, the sSFR profiles for all galaxies and for central galaxies are again similar. 
In contrast to the Illustris sample, the distribution of sSFR in EAGLE is significantly flatter for more massive galaxies and shows a steep decline with radius for lower mass galaxies.  Therefore, 
\emph{the EAGLE galaxy sample shows significantly higher central sSFR than the observational results from B18, for all mass bins.}

Although we do not directly compare, we note that the difference between observations and simulations persists in relation to the observational profiles of \citet{Nelson2016} and \citet{Ellison2018}.
\emph{The sSFR profiles of both EAGLE and Illustris suggest outside-in quenching, as opposed to the inside-out quenching often found in observations \citep[e.g.][]{Belfiore18, Nelson2016,Ellison2018}.}

\section{Discussion}\label{discussion}
 
We have identified a dramatic difference between the sSFR density profiles of Green Valley galaxies in the EAGLE and Illustris simulations and those in B18. In this section we briefly mention possible causes of this failure: the feedback prescriptions and resolution.

In the Illustris AGN feedback model, the dominant mechanism to quench galaxies is the low-accretion rate radio-mode feedback, modeled in the form of hot bubbles released in the CGM \citep{sijacki2007, vogelsberger2013}.

In EAGLE on the other hand the thermal energy from AGN feedback is deposited stochastically close to the black hole, independent of the black hole accretion rate \citep{booth2009,schaye2015}.
Additionally, differences in the star formation feedback may affect the central star formation density and central stellar mass profiles, and the simulations also use different hydrodynamic methods.

A future comparison with IllustrisTNG will be interesting as its model contains updated stellar and AGN feedback recipes with respect to the Illustris simulation while maintaining the same hydrodynamical solver, showing different distributions in many global galaxy properties \citep{Pillepich2018}, in particular an improved match to the color distribution of galaxies \citep{Nelson2018}.

Resolution is extremely important to consider in simulations, especially when examining local galaxy properties.

Indeed, measuring the sSFR profiles of galaxies may exacerbate resolution effects. Because in both simulations the gravitational softening lengths are larger than the smoothing lengths for high-density gas, the stellar mass in these galaxies will be less centrally concentrated than the gas. This purely numerical effect may result in centrally peaked sSFR profiles. While this issue may deserve its own in-depth study, in this letter we can address it in two ways.  First, we verify our results using radial bins of 0.5Rhalf, only including galaxies with Rhalf~$> 8$~kpc, so all radial bins are resolved by ${\ge}5.6$ softening lengths in order to account for the spline kernel.  Although this reduces our sample size, particularly in EAGLE, our results do not qualitatively change. 

Second, in Figure~\ref{figs:SFR} we plot the SFR volume-density profiles for the central galaxies in Illustris and EAGLE. In all of our galaxy subsets, the SFR volume-density is centrally concentrated. In Illustris (top panels), the central SFR of GV galaxies is always depressed by a smaller factor than the SFR in the outer regions (and in fact is not depressed in the center of $10 < \textrm{log}(M_*/M_{\sun}) < 11$ galaxies). 

There is generally more central SFR depression in the EAGLE GV sample, but for all galaxies $\textrm{log}(M_*/M_{\sun}) < 11$ the outer SFR is more depressed than the inner SFR, in agreement with Illustris. Only the most massive GV galaxies in EAGLE show more SFR depression within 1.5Rhalf than in the outskirts, however, for this mass bin the SF SFR radial profile is much steeper than for lower mass galaxies. 
 
Therefore the differences between the sSFR profiles in Star-Forming and GV galaxies are largely driven by the SFR profiles.  We verified that the biweight stellar mass profiles for EAGLE GV galaxies are very similar to the SF sample (not shown).  Illustris GV galaxies have either a similar mass profile to the SF sample, or are more massive in the center than at $1$Rhalf (for $\textrm{log}(M_*/M_{\sun}) > 10.5$).

Furthermore, when using the star formation rates based on the ages and birth masses of star particles (averaged over $50$~Myr or $100$~Myr), the SFR profiles of Figure~\ref{figs:SFR} essentially do not change, and neither do the sSFR profiles of Figures~\ref{figs:radIllustris} and~\ref{figs:radEAGLE}. Due to the star particle-based SFRs not recording the lowest SFRs on $50$ or $100$~Myr timescales, both the size of the galaxy sample and the extent of the star formation profiles decrease somewhat when using star particle-based SFRs.

The lowest mass bins in both simulations may also suffer from resolution effects due to a low number of gas or star particles or cells. However, only for the lowest three mass bins in the EAGLE GV sample do galaxies have less than ${\sim}100$ gas resolving elements in a number of radial bins.

Furthermore, we explore the sSFR profiles for galaxies from higher resolution smaller boxes of the EAGLE simulation suite (RefL0025N0752 and RecalL0025N0752). Although the number of galaxies is very small, the 16 central galaxies in the Green Valley (all $10.0 < \textrm{log}(M_*/M_{\sun}) < 11$) show flat profiles, showing less centrally concentrated sSFR compared to the large reference box.  However, this is still discrepant with the significant central SFR deficit of B18.
 
We have shown that the distribution of sSFR provides additional constraints on galaxy formation models, and point out that both feedback prescriptions and resolution may be important. The distribution of other galaxy properties may also provide important constraints on models, for example, stellar mass profiles \citep{Nelson2016} and gas profiles \citep{Lin2017}.

Finally we note that because SFR and stellar mass are directly computed quantities from simulations, we compare sSFR with B18. However, the sSFR profiles in B18 are based on the H$\alpha$ flux, corrected for dust attenuation using the Balmer decrement and with a correction from low-ionisation emission-line regions (LIERs). B18 only include spaxels with high S/N to ensure reliable extinction correction, but they do note that the central sSFR depends somewhat on their correction for the large fraction of LIER emission. However, they show that EW(H$\alpha$) can serve as a cross-check against dust extinction corrections and that even without LIER corrections EW(H$\alpha$) profiles show a decrement in central regions.  

\section{Conclusions}\label{conclusions}

In this paper, we identify a fundamental mismatch between the radial sSFR profiles of galaxies in the Illustris and EAGLE simulations and observations.  Specifically, in comparison to \citet{Belfiore18}, the sSFR of simulated Green Valley galaxies is too centrally concentrated.  Simulated galaxies seem to quench outside-in instead of inside-out.

We argue that sSFR profiles should be an important test of simulations, 
in addition to galaxy-wide measures like the relation between total SFR and stellar mass.  Using the sSFR profiles we also find differences between Illustris and EAGLE, likely due to differences in hydrodynamical solvers and feedback prescriptions. Differences between the GV populations in the simulations, however, are dwarfed by the dramatic difference between the sSFR profiles of both simulations and the B18 observations.
As large-scale simulations can increasingly be used to study local galaxy properties, we need to understand how these properties compare to resolved observations.

\acknowledgements
We thank 
Rachel~Somerville, Shy~Genel, and the referee
for valuable comments and suggestions. We are grateful to Francesco~Belfiore for sharing his data.
We thank the Illustris collaboration and the Virgo Consortium for
making their simulation data publicly available. The EAGLE simulations were performed using the DiRAC-2 facility at Durham, managed by the ICC, and the PRACE facility Curie based in France at TGCC, CEA, Bruy\`{e}res-le-Ch\^{a}tel.
This work was initiated as a project for the Kavli Summer Program in Astrophysics held at the Center for Computational Astrophysics of the Flatiron Institute in 2018. The program was co-funded by the Kavli Foundation and the Simons Foundation. We thank them for their generous support.
This research made use of NumPy \citep{numpy}, matplotlib, \citep{matplotlib}, Astropy, \citep{astropy}, and the corner visualization module \citep{corner}.
The Flatiron Institute is supported by the Simons Foundation.

\end{document}